\documentclass[conference]{IEEEtran}
\usepackage{cite}
\usepackage{graphicx}
\usepackage{psfrag}
\usepackage{subfigure}
\usepackage{url}
\usepackage{amsmath}
\usepackage{array}
\usepackage{amssymb}
\usepackage{amsfonts}
\newtheorem{proposition}{Proposition}

\newtheorem{corollary}{Corollary}
\newtheorem{definition}{Definition}
\newtheorem{theorem}{Theorem}
\newtheorem{lemma}{Lemma}
\newtheorem{example}{Example}
\newtheorem{observation}{Observation}
\title{Convolutional Codes for Network-Error Correction}
\author{
\authorblockN{K.~Prasad and B.~Sundar Rajan}
\authorblockA{Dept. of ECE, IISc, Bangalore 560012, India\\
Email: \{prasadk5,bsrajan\}@ece.iisc.ernet.in\\}
}
\date{\today}
\begin{document}
\maketitle
\thispagestyle{empty}
\begin{abstract}
In this work, we introduce convolutional codes  for network-error correction in the context of  coherent network coding. We give a construction of convolutional codes that correct a given set of error patterns, as long as consecutive errors are separated by a certain interval. We also give some bounds on the field size and the number of errors that can get corrected in a certain interval. Compared to previous network error correction schemes, using convolutional codes is seen to have advantages in field size and decoding technique. Some examples are discussed which illustrate the several possible situations that arise in this context.
\end{abstract}
\section{Introduction}
\label{sec1}
Network coding was introduced in \cite{ACLY} as a means to improve the rate of transmission in networks, and often achieve capacity in the case of single source networks. Linear network coding was introduced in \cite{CLY}. network-error correction, which involved a trade-off between the rate of transmission and the number of correctable network-edge errors, was introduced in \cite{YeC} as an extension of classical error correction to a general network setting. Along with subsequent works \cite{Zha} and \cite{YaY}, this generalized the classical notions of the Hamming weight, Hamming distance, minimum distance and various classical error control coding bounds to their network counterparts. An algebraic formulation of network coding was discussed in \cite{KoM} for both instantaneous networks and networks with delays. In all of these works, it is assumed that the sinks and the source know the network topology and the network code, which is referred to as \textit{coherent network coding}.

Random network coding, introduced in \cite{HMKKESL} presented a distributed network coding scheme where nodes independently chose random coefficients (from a finite field) for the linear mixing of their inputs. Subspace codes and rank metric codes were constructed for the setting of random network coding in \cite{KoK} and \cite{SKK}. 

Convolutional network codes were discussed in \cite{ErF,CLYZ,LiY} and a connection between network coding and convolutional coding was analyzed in \cite{FrS}. In this work, convolutional coding is introduced to achieve network-error correction. We assume an acyclic, single source, instantaneous (delay-free) network with coherent linear network coding to multicast information to several sinks. 

We define \textit{a network use} as a single usage of all the edges of the network to multicast utmost min-cut number of symbols to each of the sinks. An \textit{error pattern} is a subset of the set of edges of the network which are in error. It is seen that when the source implements a convolutional code to send information into the network, every sink sees a different convolutional code. We  address the following problem. 

\textit{Given an acyclic, delay-free, single-source network with a linear multicast network code, and a set of error patterns $\Phi$, how to design a convolutional code at the source which shall correct the errors corresponding to the error patterns in $\Phi$, as long as consecutive errors are separated by a certain number of network uses?}

The main contributions of this paper are as follows.
\begin{itemize}
\item For networks with a specified network code, convolutional codes have been employed to achieve network-error correction for the first time.
\item An explicit convolutional code construction (for the network with a given network code) that corrects a given pattern of network-errors (provided that the occurrence of consecutive errors are separated by certain number of network uses) is given.
\item The convolutional codes constructed in this paper are found to offer certain advantages in field size and decoding over the previous approaches of block network-error correction codes (BNECCs) of \cite{YaY} for network error correction. 
\item Some bounds are derived on the minimum field size required, and on the minimum number of network uses that two error events must be separated by in order that they get corrected. 
\end{itemize}

The rest of the paper is organized as follows. Section \ref{sec2} gives a primer on convolutional codes and MDS convolutional codes. In Section \ref{sec3}, we discuss the general network coding set-up and network-errors. In Section \ref{sec4}, we give a construction for a input convolutional code which shall correct errors corresponding to a given set of error patterns. In Section  \ref{sec5}, we give some examples for this construction. In Section \ref{sec5.5}, we compare the advantages and disadvantages of our network error correcting codes with that of \cite{YaY}. Finally, a short discussion on the construction of Section \ref{sec4} constitutes  Section \ref{sec6} along with several directions for further research.   

\section{Convolutional codes-Basic Results}
\label{sec2}
In this section, we review the basic concepts related to convolutional codes, used extensively throughout the rest of the paper. For $q,$ power of a prime, let $\mathbb{F}_q$ denote the finite field with $q$ elements.

For a convolutional code, the \textit{information sequence} $\boldsymbol{u} = \left[\boldsymbol{u}_0,\boldsymbol{u}_1,...,\boldsymbol{u}_t\right](\boldsymbol{u}_i\in\mathbb{F}_q^b)$ and the \textit{codeword sequence} (output sequence) $\boldsymbol{v} = \left[\boldsymbol{v}_0,\boldsymbol{v}_1,...,\boldsymbol{v}_t\right]\left(\boldsymbol{v}_i\in\mathbb{F}_q^c\right)$ can be represented in terms of the delay parameter $z$ as 	
\begin{eqnarray*}
\boldsymbol{u}(z)=\sum_{i=0}^t \boldsymbol{u}_i z^i ~~~ \mbox{  and  }~~~
\boldsymbol{v}(z)=\sum_{i=0}^t \boldsymbol{v}_i z^i
\end{eqnarray*}
\begin{definition}[\cite{JoZ}]
A \textit{convolutional code}, ${\cal C}$ of rate $~b/c~(b~<~c)$ is defined as 
\[
{\cal C} = \left\{ \boldsymbol{v}(z)\in\mathbb{F}_q^{c}[[z]]\text{ }|\text{ } \boldsymbol{v}(z)=\boldsymbol{u}(z)G(z) \right\}
\] 
where $G(z)$ is a $b \times c$  \textit{generator matrix} with entries from $\mathbb{F}_q(z)$(the field of rationals functions over $\mathbb{F}_q$) and rank $b$ over $\mathbb{F}_q(z)$, and $\boldsymbol{v}(z)$ being the code sequence arising from the information sequence, $\boldsymbol{u}(z)\in\mathbb{F}_q^{b}[[z]]$, the set of all $b$-tuples with elements from the formal power series ring $\mathbb{F}_q[[z]]$ over $\mathbb{F}_q.$
\end{definition}

Two generator matrices are said to be \textit{equivalent} if they encode the same convolutional code. A \textit{polynomial generator matrix}\cite{JoZ} for a convolutional code $\cal C$ is a generator matrix for $\cal C$ with all its entries from $\mathbb{F}_q[z]$, the ring of polynomials over $\mathbb{F}_q.$ It is known that every convolutional code has a polynomial generator matrix \cite{JoZ}. Also, a generator matrix for a convolutional code is \textit{catastrophic}\cite{JoZ} if there exists an information sequence with infinitely many non-zero components, that results in a codeword with only finitely many non-zero components. For a polynomial generator matrix $G(z)$, let $g_{ij}(z)$ be the element of $G(z)$ in the $i^{th}$ row and the $j^{th}$ column, and $\nu_i:=\max_{j} deg(g_{ij}(z))$ be the $i^{th}$ \textit{row degree} of $G(z)$. Let $\delta: = \sum_{i=1}^{b}\nu_i$ be the \textit{degree} of $G(z).$
\begin{definition}[\cite{JoZ} ]
A polynomial generator matrix is called \textit{basic} if it has a polynomial right inverse. It is called \textit{minimal} if its degree $\delta$ is minimum among all generator matrices of $\cal C$.
\end{definition}

Forney in \cite{For} showed that the ordered set $\left\{\nu_{1},\nu_{2},...,\nu_{b}\right\}$ of row degrees (indices) is the same for all minimal basic generator matrices of $\cal C$ (which are all equivalent to one another). Therefore the ordered row degrees and the degree $\delta$ can be defined for a convolutional code $\cal C.$ A rate $b/c$ convolutional code with degree $\delta$ will henceforth be referred to as a $(c,b,\delta)$ code. Also, any minimal basic generator matrix for a convolutional code is non-catastrophic. 

\begin{definition}[\cite{JoZ} ]
A \textit{convolutional encoder} is a physical realization of a generator matrix by a linear sequential circuit. Two encoders are said to be \textit{equivalent encoders} if they encode the same code. A \textit{minimal encoder} is an encoder with minimal delay elements among all equivalent encoders.
\end{definition}
\begin{definition}[\cite{JoZ}]
The \textit{free distance} of the convolutional code $\cal C$ is given as 
\[
d_{free}({\cal C})=min\left\{w_H(\boldsymbol{v}(z))|\boldsymbol{v}(z)\in{\cal C},\boldsymbol{v}(z)\neq 0\right\}
\] 
where $w_H$ indicates the Hamming weight over $\mathbb{F}_q.$
\end{definition}
\subsection{MDS convolutional codes}

In this subsection, we discuss some results on the existence and construction of Maximum Distance Separable (MDS) convolutional codes. In Subsection \ref{sec4e}, we use these results to obtain some bounds on the field size and the error correcting capabilities of such MDS convolutional codes when they are used for network-error correction. The following bound on the free distance, and the existence of codes meeting the bound, called MDS convolutional codes, was proved in \cite{RoS}. 
\begin{theorem}[\cite{RoS}]
\label{GenSingBound}
For every base field $\mathbb{F}$ and every rate $k/n$ convolutional code $\cal C$ of degree $\delta$, the free distance is bounded as 
\[
d_{free}({\cal C})\leq(n-k)(\left\lfloor \delta / k \right\rfloor + 1) + \delta + 1.
\]
\end{theorem}

Theorem \ref{GenSingBound} is known as the \textit{generalized Singleton bound}.	
\begin{theorem}[\cite{RoS}]For any positive integers $k<n$, $\delta$ and for any prime $p$ there exists a field $\mathbb{F}_q$ of characteristic $p$, and a rate $k/n$ convolutional code $\cal C$ of degree $\delta$ over $\mathbb{F}_q$, whose free distance meets the generalized Singleton bound.
\end{theorem}

A method of constructing MDS convolutional codes based on the connection between quasi-cyclic codes and convolutional codes was given in \cite{RLS}. It is known \cite{RLS} that the field size $q$ required for a $(n,k,\delta)$ MDS convolutional code ${\cal C}$ in the construction in \cite{RLS} should be a prime power such that 
\begin{equation}
\label{fieldsizeconv}
n|(q-1)\text{ and } q\geq\delta\frac{n^2}{k(n-k)}+2.
\end{equation}

\section{Convolutional Codes for network-error Correction - Problem Formulation}
\label{sec3}
\subsection{Network model}
We consider only acyclic networks in this paper the model for which is as in \cite{CLYZ}. An acyclic network can be represented as a acyclic directed multi-graph ${\cal G}$ = ($\cal V,\cal E$) where $\cal V$ is the set of all vertices and $\cal E$ is the set of all edges in the network. 

We assume that every edge in the directed multi-graph representing the network has unit \emph{capacity} (can carry utmost one symbol from $\mathbb{F}_q$). Network links with capacities greater than unit are modeled as parallel edges. The network is assumed to be instantaneous, i.e, all nodes process the same \emph {generation} (the set of symbols generated at the source at a particular time instant) of input symbols to the network in a given coding order (ancestral order \cite{CLYZ}).

Let $s\in\cal V$ be the source node and $\cal T$ be the set of all receivers. Let $n_{_T}$ be the unicast capacity for a sink node  $T\in{\cal T}$ i.e the maximum number of edge-disjoint paths from $s$ to $T$. Then $n = \min_{T\in{\cal T}}n_{_T}$ is the max-flow min-cut capacity of the multicast connection. 
\subsection{Network code}
We follow  \cite{KoM} in describing the network code. For each node $v\in{\cal V}$, let the set of all incoming edges be denoted by $\Gamma_I(v)$. Then $|\Gamma_I(v)|=\delta_I(v)$ is the in-degree of $v$. Similarly the set of all outgoing edges is defined by $\Gamma_O(v)$, and the out-degree of the node $v$ is given by $|\Gamma_O(v)|=\delta_O(v)$.  For any $e \in {\cal E}$ and $v \in {\cal V}$, let $head(e)=v$, if $v$ is such that $e \in \Gamma_I(v)$. Similarly, let $tail(e)=v$, if $v$ is such that $e \in \Gamma_O(v)$. We will assume an ancestral ordering on ${\cal E}$ of the acyclic graph ${\cal G}$.

The network code can be defined by the local kernel matrices of size $\delta_I(v)\times\delta_O(v)$ for each node $v\in{\cal V}$ with entries from $\mathbb{F}_q$. The global encoding kernels for each edge can be recursively calculated from these local kernels.

The network transfer matrix, which governs the input-output relationship in the network, is defined as given in \cite{KoM}. Towards this end, the matrices $A$,$K$,and $B^T$(for every sink $T\in {\cal T}$ are defined as follows:\newline
The entries of the $n \times |{\cal E}|$ matrix $A$ are defined as
\[
A_{i,j}=\left\{
\begin{array}{cc}
\alpha_{i,e_j} & \text{   if } e_j \in \Gamma_{O}(s)\\
0  & \text{ otherwise}
\end{array}
\right. 
\]
where $\alpha_{i,e_j} \in \mathbb{F}_q$ is the local encoding kernel coefficient at the source coupling input $i$ with edge $e_j \in \Gamma_O(s)$.\newline 
The entries of the $|{\cal E}| \times |{\cal E}|$ matrix $K$ are defined as
\[
K_{i,j}=\left\{
\begin{array}{cc}
\beta_{i,j} & \text{   if } head(e_i) = tail(e_j) \\
0  & \text{ otherwise} 
\end{array}
\right. 
\]
where the set of $\beta_{i,j} \in \mathbb{F}_q$ is the local encoding kernel coefficient  between $e_i$ and $e_j$, at the node $v=head(e_i) = tail(e_j)$.\newline
For every sink $T \in {\cal T}$, the entries of the $|{\cal E}| \times n$ matrix $B^T$ are defined as  
\[
B^T_{i,j}=\left\{
\begin{array}{cc}
\epsilon_{e_j,i} & \text{   if } e_j \in \Gamma_{I}(T)\\
0  & \text{ otherwise} \end{array} 
\right. 
 \]
where all $\epsilon_{e_j,i} \in \mathbb{F}_q$.

For instantaneous networks, we have 
\begin{eqnarray*}
F : = (I-K)^{-1} 
\end{eqnarray*}
where $I$ is the $|{\cal E}| \times |{\cal E}|$ identity matrix.  Now we have the following:
\begin{definition}[\cite{KoM}]
\label{nettransfermatrix}
\textit{The network transfer matrix}, $M_{T}$, corresponding to a sink node ${T} \in \cal T$ is a full rank $n \times n$ matrix defined as $~~~M_{T}:=AFB^{T}=AF_{T}.$ 
\end{definition} 	

Definition \ref{nettransfermatrix} implies that if $\boldsymbol{x} \in \mathbb{F}_q^n$ is the input to the instantaneous network at any particular instant, then at any particular sink $T \in \cal T$, we have the output, $\boldsymbol{y} \in \mathbb{F}_q^n$, at the same instant, to be $\boldsymbol{y} = \boldsymbol{x}M_T$.
\subsection{Convolutional codes for networks}
Assuming that a $n$-dimensional linear network code multicast has been implemented in the network, we define the following terms-
\begin{definition}
An \textit{input convolutional code}, ${\cal C}_s$ is a convolutional code of rate $~k/n (k < n)$ with a \textit{input generator matrix }$G_{I}(z)$ implemented at the source of the network.
\end{definition}
\begin{definition}
The \textit{output convolutional code} ${\cal C}_T$, corresponding to a sink node ${T} \in \cal T$ is the $~k/n (k < n)$ convolutional code generated by the \textit{output generator matrix } $G_{O,{T}}(z)$ which is given as $~~~G_{O,{T}}(z) = G_I(z)M_{T}$, with $M_T$ being the full rank network transfer matrix corresponding to a $n$-dimensional network code.
\end{definition}
\begin{example} 
\label{exm1}
\begin{figure}[htbp]
\centering
\includegraphics[totalheight=2.2in,width=3in]{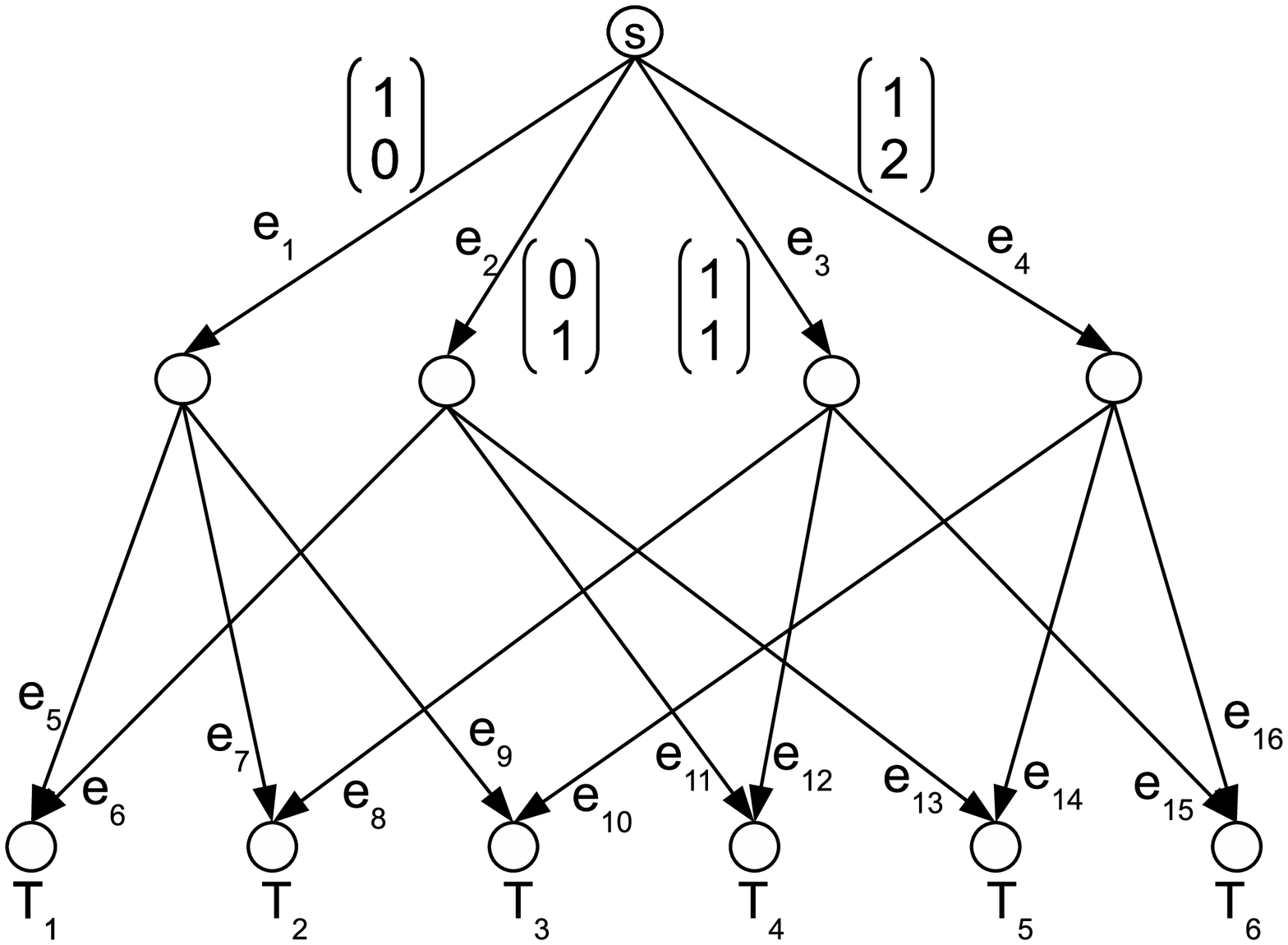}
\caption{$_4C_2$ combination network over a ternary field. The global kernels of the edges coming from the source are indicated. All the intermediate nodes have local kernels unity.}	
\label{fig:4C2Network}	
\end{figure}
Consider the $_4C_2$ combination $\mathbb{F}_3$ network as shown in Fig. \ref{fig:4C2Network}. For this network, let the input convolutional code over $\mathbb{F}_3[z]$ be generated by $G_I(z) = \left[1 + z^2\text{ }\text{ }1+z+z^2\right].$ 
The network transfer matrices at each sink and their corresponding output convolutional matrices are calculated and tabulated in Table \ref{tab1}. 
\begin{table}[htbp]
\centering
\caption{$_4C_2-$ $\mathbb{F}_3$ network for the input convolutional code $G_I(z) = \left[1 + z^2\text{ }\text{ }1+z+z^2\right].$}
\begin{tabular}{|c|c|l|}\hline
\textbf{Sink} & \textbf{Network transfer} & \textbf{Output convolutional code}\\
              & \textbf{matrix}           &                                   \\ \hline
$T_1$ & $ M_{T_1}= \left( \begin{array}{cc}1 & 0 \\0  & 1 \end{array} \right)$ & $G_{O,T_1}(z) =[1+z^2\text{ }\text{ }1+z+z^2]$\\
\hline
$T_2$ & $M_{T_2}=\left( \begin{array}{cc}1 & 1 \\0  & 1 \end{array} \right)$ & $G_{O,T_2}(z) =[1+z^2\text{ }\text{ }2+z+2z^2]$\\
\hline
$T_3$ & $M_{T_3}=\left( \begin{array}{cc}1 & 1 \\0  & 2 \end{array} \right)$ & $G_{O,T_3}(z) =[1+z^2\text{ }\text{ }2z]$\\
\hline
$T_4$ & $M_{T_4}=\left( \begin{array}{cc}0 & 1 \\1  & 1 \end{array} \right)$ & $G_{O,T_4}(z) =[1+z+z^2\text{ }\text{ }2+z+2z^2]$\\
\hline
$T_5$ & $M_{T_5}=\left( \begin{array}{cc}0 & 1 \\1  & 2 \end{array} \right)$ & $G_{O,T_5}(z) =[1+z+z^2\text{ }\text{ }2z]$\\
\hline
$T_6$ & $M_{T_6}=\left( \begin{array}{cc}1 & 1 \\1  & 2 \end{array} \right)$ & $G_{O,T_6}(z) =[2+z+2z^2\text{ }\text{ }2z]$\\
\hline
\end{tabular}
\label{tab1}
\end{table}
\end{example}
\begin{figure*}
[htbp]
\centering
\includegraphics[totalheight=3in,width=5.6in]{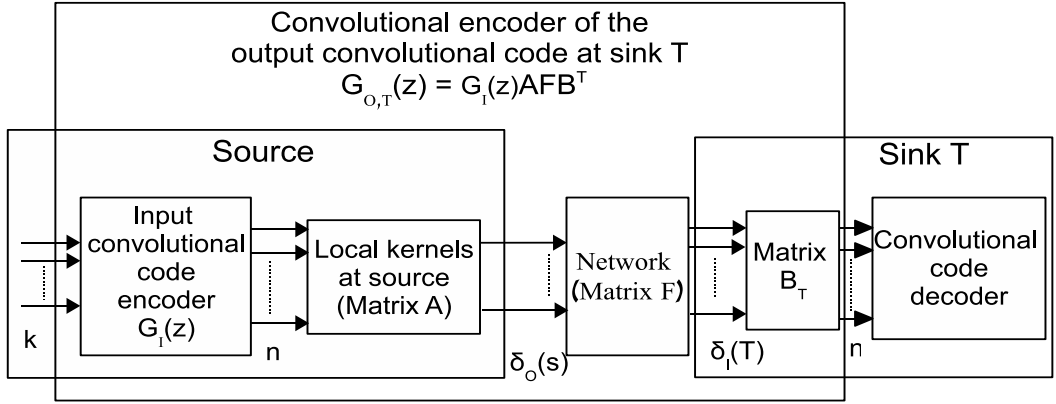}
\caption{A network with a input convolutional code and a network code}	
\label{fig:netcodewithcc}
~\\ \hrule
\end{figure*}

Thus, as can be seen from the Example \ref{exm1}, the source node implements a convolutional code, and maps the encoded symbols into its outgoing symbols. The network maps these symbols from the source to symbols at the receivers. Each of the receivers hence sees a different convolutional code which might have different distance properties and different degrees ($\delta$). Fig. \ref{fig:netcodewithcc} illustrates the entire system for a particular sink. 

\subsection{Network-errors}
An \textit{error pattern} $\rho,$ as stated previously, is a subset of ${\cal E}$ which indicates the edges of the network in error. An \textit{error vector} $\boldsymbol{w}$ is a $1\times |{\cal E}|$ vector which indicates the error occurred at each edge. An error vector is said to match an error pattern $(\text{i.e }\boldsymbol{w} \in \rho)$ if all non-zero components of $\boldsymbol{w}$ occur only on the edges in $\rho$. An \textit{error pattern set} $\Phi$ is a collection of subsets of ${\cal E}$, each of which is an error pattern. Therefore we have the formulation as follows.

Let $\boldsymbol{x} \in \mathbb{F}_q^n$ be the input to the network at any particular time instant, and let $\boldsymbol{w} \in F_q^{|{\cal E}|}$ be the error vector corresponding to the network-errors that occurred in the same particular instant. Then, the output vector, $\boldsymbol{y} \in \mathbb{F}_q^n$, at that instant at any particular sink $T \in \cal T$ can be expressed as 
\[
\boldsymbol{y} = \boldsymbol{x}M_T + \boldsymbol{w}F_T
\]
\section{Convolutional Codes for network-error Correction - Code Construction and Capability}
\label{sec4}
\subsection{Bounded distance decoding of convolutional codes}
\label{sec4a}
In this section, we briefly discuss and give some results regarding the bounded distance decoding of convolutional codes, reinterpreting results from \cite{JeR} for our context.

For the convolutional encoder with $c$ encoded output symbols and $b$ input (information) symbols, starting at some state in the trellis, we shall denote every such $c$ output symbol durations as \textit{a segment} of the trellis of the convolutional code. Each segment can be identified by an integer, which is zero at the start of transmission and  incremented by $1$ for every $c$ output symbols henceforth.  

Let $\cal C$ be a rate $b/c$ convolutional code with a  generator matrix $G(z).$ Then corresponding to the information sequences $\boldsymbol{u}_0,\boldsymbol{u}_1,.. (\boldsymbol{u}_i \in \mathbb{F}_q^b)$ and the code sequence $\boldsymbol{v}_0,\boldsymbol{v}_1,... (\boldsymbol{v}_i \in \mathbb{F}_q^c)$, we can associate an encoder state sequence $\boldsymbol{\sigma}_0,\boldsymbol{\sigma}_1,. . $, where 	$\boldsymbol{\sigma}_t$ indicates the content of the delay elements in the encoder at a time $t.$We define the set of $j$ output symbols as $\boldsymbol{v}_{[0,j)}:=\left[\boldsymbol{v}_0,\boldsymbol{v}_1,. . . ,\boldsymbol{v}_{j-1}\right].$ We define the set $S_{d_{free}}$ consisting of all possible truncated code sequences $\boldsymbol{v}_{[0,j)}$ $\forall$ $j$ of weight less than $d_{free}({\cal C})$ that start in the zero state as follows:
\begin{eqnarray*}
S_{d_{free}}:=\left\{\boldsymbol{v}_{[0,j)} \mid w_H\left(\boldsymbol{v}_{[0,j)}\right) < d_{free}({\cal C}),  \boldsymbol{\sigma}_0=\boldsymbol{0}, ~ \forall~ j>0 \right\}
\end{eqnarray*}
where $w_H$ indicates the Hamming weight over $\mathbb{F}_q.$ Clearly the definition of $S_{d_{free}}$ excludes the possibility of a zero state in between (in the event of which $w_H\left(\boldsymbol{v}_{[0,j)}\right) \geq d_{free}({\cal C})$), i.e, $ \boldsymbol{\sigma}_t \neq \boldsymbol{0} \text{ for any t such that } 0 < t \leq j.$ We have that the set $S_{d_{free}}$ is invariant among the set of minimal convolutional encoders. We now define
\[
T_{d_{free}}({\cal C}):=\max_{\boldsymbol{v}_{[0,j)} \in S_{d_{free}}}j+1
\]
which thereby can be considered as a code property because of the fact that $S_{d_{free}}$ is invariant among minimal encoders. Then, we have the following proposition:
\begin{proposition}
\label{minweighttime}
The minimum Hamming weight trellis decoding algorithm can correct all error sequences which have the property that the Hamming weight of the error sequence in any consecutive $T_{d_{free}}({\cal C})$ segments is utmost $\left\lfloor \frac{d_{free}({\cal C})-1}{2} \right\rfloor$.
\end{proposition}
\begin{proof}
Without loss of generality, let $\boldsymbol{\sigma}_t$ be a correct state (according to the transmitted sequence) at some depth $t$ in the path traced by the received sequence on the trellis of the code and let us assume that all the errors before $t$ have been corrected. 

Now consider the window from $t$ to $t+T_{d_{free}}({\cal C})$, consisting of $T_{d_{free}}({\cal C})$ segments. In this window, the Hamming weight of the error sequence is utmost $\left\lfloor \frac{d_{free}({\cal C})-1}{2} \right\rfloor$. However, by the definition of $T_{d_{free}}({\cal C})$, the distance between the correct path and every other path of length $T_{d_{free}}({\cal C})$ starting from the state $\boldsymbol{\sigma}_t$ is at least $d_{free}({\cal C})$. Therefore, in this window, the error sequence can be corrected. 

Now using $\boldsymbol{\sigma}_{t+T_{d_{free}}({\cal C})}$ at depth $t+T_{d_{free}}({\cal C})$ as our new correct starting state, we can repeat the same argument thus proving that the entire error sequence is correctable. 
\end{proof}
\subsection{Construction}
\label{construction} 
For the given network with a single source that has to multicast information to a set of sinks,  $n$ being min-cut of the multicast connections, a $n$-dimensional network code in place over a sufficiently large field $F_{q}$ (for which we provide a bound in Subsection \ref{sec4e}) of characteristic $p$, we provide a construction for a convolutional code for correcting errors with patterns in a given error pattern set. This is the main contribution of this work. The construction is as follows. 
\begin{enumerate}
\item Let $M_T=AF_{T}$ be the $n\times n$ network transfer matrix from the source to any particular sink $T\in {\cal T}$. Let $\Phi$ be the error pattern set given. Then we compute the following sets.
\item Let the set of all error vectors having their error pattern in $\Phi$ be 
\[
{\cal W}_{\Phi}=\bigcup_{\rho \in \Phi}\left\{\boldsymbol{w}=(w_1,w_2,...,w_{|{\cal E}|}) \in \mathbb{F}_{q}^{|{\cal E}|}\text{ }|\text{ }\boldsymbol{w} \in \rho\right\}.
\]
\item Let 
\[
{\cal W}_{T}:= \left\{\boldsymbol{w}F_T\text{ }|\text{ }\boldsymbol{w}\in{\cal W}_{\Phi}\right\}
\] 
be computed for each sink $T$. This is nothing but the set of n-length resultant vectors at the sink $T$ due to errors in the given error patterns $\rho \in \Phi$. 
\item Let 
\[
{\cal W}_s:=\bigcup_{T\in{\cal T}} \left\{\boldsymbol{w}_{_T}M_T^{-1}\text{ }|\text{ }\boldsymbol{w}_{_T}\in{\cal W}_{T}\right\} 
\] 
be computed. This is the set of all $n$ length input vectors to the network that would result in the set of output vectors given by ${\cal W}_{T}$ at sink $T$, for each sink $T$. 
\item Given a vector $\boldsymbol{y} \in \mathbb{F}_q^m$ (for some positive integer $m$), let $w_H(\boldsymbol{y})$ denote the Hamming weight of $\boldsymbol{y}$, i.e., the number of non-zero elements of $\boldsymbol{y}$. Let 
\begin{equation}
\label{ts}
t_s = \max_{\boldsymbol{w}_s \in {\cal W}_s}w_H(\boldsymbol{w}_s).
\end{equation}
\item Choose an input convolutional code ${\cal C}_s$ with free distance at least $2t_s+1$. 
\end{enumerate}
\subsection{Decoding}
\label{decoding}
Let $G_I(z)$ be the $k \times n$ generator matrix of the input convolutional code, ${\cal C}_s$, obtained from the given construction. Let $G_{O,{T}}(z) = G_I(z)M_{T}$ be the generator matrix of the output convolutional code, ${\cal C}_{T}$, at sink $T \in {\cal T}$, with $M_T$ being its network transfer matrix. Each sink can choose between two decoding methods based on the free distance ($d_{free}({\cal C}_{T})$) and $T_{d_{free}}({\cal C}_T)$ of its output convolutional code as follows:

\textit{Case-A:} This case is applicable in the event of both of the following two conditions are satisfied. 
\begin{equation}\label{decodAcond1} 
d_{free}({{\cal C}_T}) \geq 2 \left( \max_{\boldsymbol{w}_{_T} \in {\cal W}_T} w_H(\boldsymbol{w}_{_T}) \right)+1 
\end{equation} 
and  
\begin{equation}\label{decodAcond2} 
T_{d_{free}}({\cal C}_s) \geq T_{d_{free}}({\cal C}_T).
\end{equation}
In this case, the sink $T$ performs minimum distance decoding directly on the trellis of the output convolutional code, ${\cal C}_{T}$.

\textit{Case-B:} This case is applicable if either of the following two conditions are satisfied.  
\[
d_{free}({{\cal C}_T}) < 2 \left(\max_{\boldsymbol{w}_{_T} \in {\cal W}_T}w_H(\boldsymbol{w}_{_T}) \right)+1
\]
or 
\[
T_{d_{free}}({\cal C}_s) < T_{d_{free}}({\cal C}_T).
\]
This method involves additional processing at the sink, i.e, matrix multiplication. We have the following formulation at the sink $T$. Let 
\begin{eqnarray*} {\left[v_{1}'(z) \text{ }\text{ } v_{2}'(z) \text{ }\text{ } ... \text{ }\text{ } v_{n}'(z)\right] =  \left[v_{1}(z)\text{ }\text{ }v_{2}(z)\text{ }\text{ } ... \text{ }\text{ }v_{n}(z)\right]  + }\\  \left[w_{1}(z)\text{ }\text{ }w_{2}(z)\text{ }\text{ }...\text{ }\text{ }w_{n}(z)\right] \end{eqnarray*} represent the output sequences at sink $T$, where 
\[
\left[v_{1}(z) \text{ }\text{ } v_{2}(z) \text{ }\text{ } ... \text{ }\text{ } v_{n}(z)\right] = \boldsymbol{u}(z)G_{O,T}(z) = \boldsymbol{u}(z)G_{I}(z)M_T 
\]
$\boldsymbol{u}(z)$ being the $k$ length vector of input sequences, and 
\[
\left[w_{1}(z) \text{ }\text{ } w_{2}(z) \text{ }\text{ } ... \text{ }\text{ }w_{n}(z)\right]
\]
represent the corresponding error sequences. Now, the output sequences are multiplied with the inverse of the network transfer matrix $M_T$, so that decoding can be done on the trellis of the input convolutional code. Hence, we have 
\begin{eqnarray*} \left[v_{1}''(z) \text{ }\text{ } v_{2}''(z) \text{ }\text{ } ... \text{ }\text{ } v_{n}''(z) \right] =  \left[v_{1}'(z) \text{ }\text{ } v_{2}'(z) \text{ }\text{ } ... \text{ }\text{ } v_{n}'(z) \right]M_T^{-1} \\
= \boldsymbol{u}(z)G_{I}(z) + \left[w_{1}(z) \text{ }\text{ } w_{2}(z) \text{ }\text{ } ... \text{ }\text{ }w_{n}(z) \right]M_T^{-1}
\\=\boldsymbol{u}(z)G_{I}(z) + \left[w_{1}'(z) \text{ }\text{ } w_{2}'(z) \text{ }\text{ } ... \text{ }\text{ }w_{n}'(z) \right] \end{eqnarray*}
where  $\boldsymbol{w'}(z) = \left[w_{1}'(z) \text{ }\text{ } w_{2}'(z) \text{ }\text{ } ... \text{ }\text{ }w_{n}'(z) \right]$ now indicate the set of modified error sequences that are to be corrected. Then the sink $T$ decodes to the minimum distance path on the trellis of the input convolutional code. 
\subsection{Error correcting capability}
\label{capability}
In this subsection we prove a main result of the paper given by Theorem \ref{maintheorem} which characterizes the error correcting capability of the code obtained via the construction of Subsection \ref{construction}. Before proving the following theorem, we recall the following observation that in every network use, $n$ encoded symbols which is equal to the number of symbols corresponding to one segment of the trellis, are to be multicast to the sinks.
\begin{theorem}
\label{maintheorem}
The code ${\cal C}_s$ resulting from the construction of Subsection \ref{construction} can correct all network-errors that have their pattern as some $\rho \in \Phi$ as long as any two consecutive network-errors are separated by $T_{d_{free}}({\cal C}_s)$ network uses.
\end{theorem}
\begin{proof}
In the event of Case-A of the decoding, the given conditions ((\ref{decodAcond1}) and (\ref{decodAcond2})) together with Proposition \ref{minweighttime} prove the given claim that errors with their error pattern in $\Phi$  will be corrected as long as no two consecutive error events occur within $T_{d_{free}}({\cal C}_s)$ network uses. 

In fact, condition (\ref{decodAcond1}) implies that network-errors with pattern in $\Phi$ will be corrected at sink $T$, as long as consecutive error events are separated by $T_{d_{free}}({\cal C}_{T})$.

Now we consider Case B of the decoding. Suppose that the set of error sequences in the formulation given, $\boldsymbol{w'}(z)$, is due to network-errors that have their pattern as some $\rho \in \Phi$, such that any two consecutive such network-errors are separated by  at least $T_{d_{free}}({\cal C}_s)$ network uses. 

Then, from (\ref{ts}), we have that the maximum Hamming weight of any error event embedded in $\boldsymbol{w'}(z)$ would be utmost $t_s$, and any two consecutive error events would be separated by $T_{d_{free}}({\cal C}_s)$ segments of the trellis of the code ${\cal C}_s$. Because of the free distance of the code chosen and along with Proposition \ref{minweighttime}, we have that such errors will get corrected when decoding on the trellis of the input convolutional code.
\end{proof}
\subsection{Bounds on the field size and $T_{d_{free}}({\cal C}_s)$}
\label{sec4e}
\subsubsection{Bound on field size}
The following theorem gives a sufficient field size for the required $(n,k)$ convolutional code to be constructed with the required free distance condition ($\geq 2t_s+1$).
\begin{theorem}
\label{fieldsizebound}
The code ${\cal C}_s$ can be constructed and used to multicast $k$ symbols to the set of sinks ${\cal T}$ along with the required error correction in the given instantaneous network with min-cut $n$ ($n>k$), if the field size $q$ is such that 
\[
n|q-1 ~~~~\text{ and } ~~~~ q > max\left\{|{\cal T}|,\frac{2n^2}{n-k}+2 \right\}.
\] 
\end{theorem}
\begin{proof}
The condition that 
\[
q>|{\cal T}|
\]
is from the known sufficient condition \cite{HMKKESL} for the existence of a linear multicast network code. \newline
For the other conditions, we first note that in the construction of Subsection \ref{construction}, $t_s\leq n$. In the worst case that $t_s=n$, we need $d_{free}({\cal C}_s) \geq 2n+1$. We have from the generalized Singleton bound:
\[
d_{free}({\cal C}_s)\leq(n-k)(\left\lfloor \delta / k \right\rfloor + 1) + \delta + 1.
\]
In order that $d_{free}({\cal C}_s)$ be at least $2n+1$, we let $\delta = 2k$, in which case the R.H.S of the inequality becomes 
\begin{eqnarray*}
(n-k)(\left\lfloor 2k / k \right\rfloor + 1) + 2k + 1 \\
= 2n+(n-k)+1 > 2n+1
\end{eqnarray*}
Thus, with $\delta = 2k$, from (\ref{fieldsizeconv}) we have that $(n,k,\delta = 2k)$ MDS convolutional code can be constructed based on \cite{RLS} if 
\[
n|q-1 \text{ and } q > \frac{2n^2}{n-k}+2.
\]
Such an MDS convolutional code the requirements in the construction ($d_{free}({\cal C}_s) \geq 2n+1$), and hence the theorem is proved. 
\end{proof}
\subsubsection{Bound on $T_{d_{free}}({\cal C}_s)$}
Towards obtaining a bound on $T_{d_{free}}({\cal C}_s)$, we first prove the following lemma. 
\begin{lemma}
\label{deltazeroes}
Let $\cal C$ be a rate $b/c$ convolutional code with degree $\delta$ and $S_{d_{free}}$ be defined as in Subsection \ref{sec4a} for a minimal encoder (a controller canonical form realization \cite{JoZ} of a minimal basic generator matrix, $G_{mb}(z)$, of $\left.{\cal C}\right)$. Then any 
\[
\boldsymbol{v}_{[0,j)} = \left[\boldsymbol{v}_0,\boldsymbol{v}_1,. . . ,\boldsymbol{v}_{j-1}\right] \in S_{d_{free}}
\]
cannot have $\delta$ zeros in $\delta$ consecutive segments, i.e, at least one of $\boldsymbol{v}_i,\boldsymbol{v}_i,...,\boldsymbol{v}_{i+\delta-1}$ is non zero $ \forall$ $0 \leq i \leq j- \delta.$
\end{lemma}
\begin{IEEEproof}
Let the ordered Forney indices (row degrees of $G_{mb}(z)$) be $\nu_1,\nu_2,. . . ,\nu_b=\nu_{max}$, and therefore $\delta$ being the sum of these indices. Then a systematic generator matrix($G_{sys}(z)$) for $\cal C$ that is equivalent to $G_{mb}(z)$ is of the form 
\begin{equation*}
G_{sys}(z)=T^{-1}(z)G_{mb}(z)
\end{equation*}
where $T(z)$ is a full rank $b \times b$ submatrix of $G_{mb}(z)$ with a delay-free determinant. We have the following observation. 
\begin{observation}
\label{degree}
The degree of $det\left(T(z)\right)$ is clearly utmost $\delta.$ Also, we have the $(i,j)^{th}$ element $t_{i,j}(z)$ of $T^{-1}(z)$ as
\[
t_{i,j}(z)= \frac{Cofactor\left(T(z)_{j,i}\right)}{det\left(T(z)\right)}
\]
where $Cofactor(T(z)_{j,i}) \in \mathbb{F}_q[z]$ is the cofactor of the $(j,i)^{th}$ element of $T(z).$ The degree of $Cofactor(T(z)_{j,i})$ is utmost $\delta - \nu_j \leq \delta - \nu_1.$ 

Let $a_{i,j}(z) \in \mathbb{F}_q(z)$ represent the $(i,j)^{th}$ element of $G_{sys}(z),$ where 
\begin{eqnarray*}
a_{i,j}(z)=\sum_{k=1}^{b}t_{i,k}(z)g_{k,j}(z)~~~~~~~~ \\
~~~~~~~= \frac{\sum_{k=1}^{b}Cofactor(T(z)_{k,i})g_{k,j}(z)}{det\left(T(z)\right)}
\end{eqnarray*}
$g_{k,j}(z)$ being $(k,j)^{th}$ element of $G_{mb}(z).$ Therefore, the element $a_{i,j}(z)$ can be expressed as
\[
a_{i,j}(z) = \frac{p_{i,j}(z)}{det\left(T(z)\right)}
\]
where the degree of $p_{i,j}(z) \in \mathbb{F}_q[z]$ is utmost $\delta + \nu_{max} - \nu_1.$ Now if we divide $p_{i,j}(z)$ by $det\left(T(z)\right)$, we have
\begin{equation}
\label{eqn4}
a_{i,j}(z)= q_{i,j}(z) + \frac{r_{i,j}(z)}{det\left(T(z)\right)}
\end{equation}
where the degree of $q_{i,j}(z) \in \mathbb{F}_q[z]$ is utmost $\nu_{max}-\nu_1$, and the degree of $r_{i,j}(z)$ is utmost $\delta - 1.$ Because every element of $G_{sys}(z)$ can be reduced to the form in (\ref{eqn4}), we can have a realization of $G_{sys}(z)$ with utmost $\delta$ memory elements for each of the $b$ inputs. Let this encoder realization be known as $E.$
\end{observation}

Now we shall prove the lemma by contradiction. Suppose there exists a codeword  $\boldsymbol{v}(z)=\left[\boldsymbol{v}_0,\boldsymbol{v}_1,...,\boldsymbol{v}_{j-2},\boldsymbol{v}_{j-1},\boldsymbol{v}_j,...\right]$ exists such that $\boldsymbol{v}_{[0,j)} = \left[\boldsymbol{v}_0,\boldsymbol{v}_1,. . . ,\boldsymbol{v}_{j-1}\right] \in S_{d_{free}}$ and $\boldsymbol{v}_i,\boldsymbol{v}_i,...,\boldsymbol{v}_{i+\delta-1}$ are all zero  for some $i$ such that $0 \leq i \leq j- \delta.$

Let $\boldsymbol{u}_s(z)$ be the information sequence which when encoded into $\boldsymbol{v}(z)$ by the systematic encoder $E.$ Because of the systematic property of $E$, we must have that $\boldsymbol{u}_i,\boldsymbol{u}_i,...,\boldsymbol{u}_{i+\delta-1}$ are also all zero. By Observation \ref{degree}, $E$ is an encoder which has utmost $\delta$ delay elements (for each input), and hence the state vector $\boldsymbol{\sigma}_{i+\delta}$ at time instant $i+\delta$ becomes zero as a result of these $\delta$ zero input vectors. Fig. \ref{fig:tdfreebound} shows the scenario we consider.
\begin{figure}[htbp]
\centering
\includegraphics[totalheight=2.66in,width=3.2in]{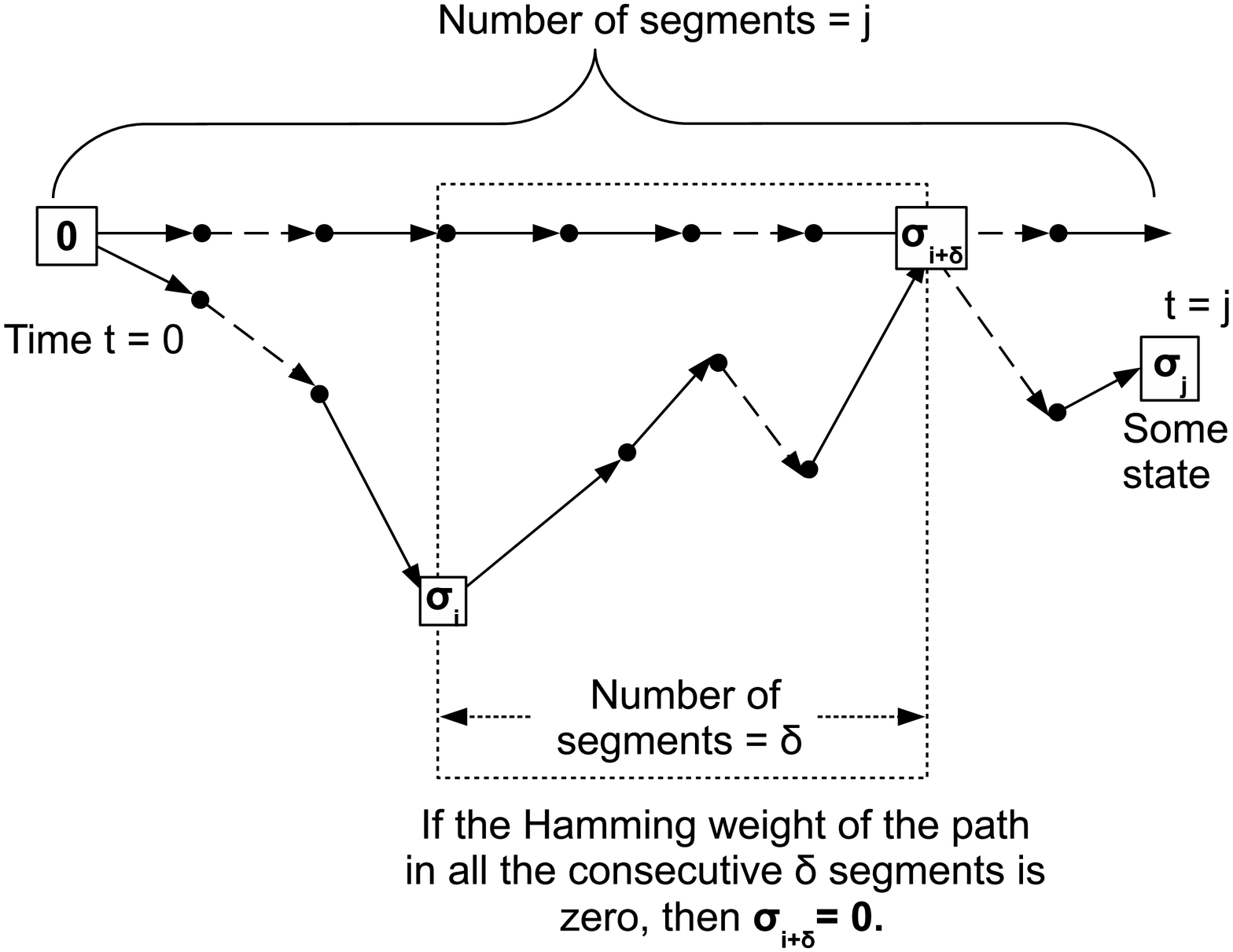}
\caption{The trellis corresponding to a systematic encoder of $\cal C$}	
\label{fig:tdfreebound}	
\end{figure}

Therefore the codeword $\boldsymbol{v}(z)$ can be written as a unique sum of two code words $\boldsymbol{v}(z)=\boldsymbol{v}'(z)+\boldsymbol{v}''(z)$, where
\[
\boldsymbol{v}'(z)=\sum_{k=1}^{i+\delta-1}\boldsymbol{v}_kz^k=\left[\boldsymbol{v}_0,...,\boldsymbol{v}_{i}=\boldsymbol{0},...,\boldsymbol{v}_{i+\delta-1}=\boldsymbol{0},\boldsymbol{0},...\right]
\]
and 
\[
\boldsymbol{v}''(z)=\sum_{k=i+\delta}\boldsymbol{v}_kz^k=\left[\boldsymbol{0},\boldsymbol{0},...,\boldsymbol{0},\boldsymbol{v}_{i+\delta},\boldsymbol{v}_{i+\delta+1},...,\boldsymbol{v}_{j},...\right]
\]
where $\boldsymbol{0}\in \mathbb{F}_q^c$ and the uniqueness of the decomposition holds with respect to the positions of the zeros indicated in the two code words $\boldsymbol{v}'(z)$ and $\boldsymbol{v}''(z).$ 

Let $\boldsymbol{u}_{mb}(z)$ be the information sequence which is encoded into $\boldsymbol{v}(z)$ by a minimal realization $E_{mb}$ of a minimal basic generator matrix $G_{mb}(z)$ (a minimal encoder). Then we have
\[
\boldsymbol{u}_{mb}(z)=\boldsymbol{u}'_{mb}(z)+\boldsymbol{u}''_{mb}(z)
\]
where $\boldsymbol{u}'_{mb}(z)$ and $\boldsymbol{u}''_{mb}(z)$ are encoded by $E_{mb}$ into $\boldsymbol{v}'(z)$ and $\boldsymbol{v}''(z)$ respectively.

By the \textit{predictable degree property} (PDP) \cite{JoZ} of minimal basic generator matrices, we have that for any polynomial code sequence $\boldsymbol{v}(z)$,
\[
deg\left(\boldsymbol{v}(z)\right)=\max_{1\leq l \leq b}\left\{deg\left(\boldsymbol{u}_{mb,l}(z)\right)+\nu_l\right\}.
\]
where $\boldsymbol{u}_{mb,l}(z) \in \mathbb{F}_q[z]$ represents the information sequence corresponding to the $l^{th}$ input, and $deg$ indicates the degree of the polynomial. Therefore, by the PDP property, we have that $deg\left(\boldsymbol{u}'_{mb}(z)\right) < i$, since  $deg\left(\boldsymbol{v}'(z)\right)<i$. 

Also, it is known that in the trellis of corresponding to a minimal realization of a minimal-basic generator matrix, there exists no non-trivial transition from the all-zero state to a non-zero state that produces a zero output. Therefore we have $deg\left(\boldsymbol{u}''_{mb}(z)\right) \geq i+\delta$, with equality being satisfied if $\boldsymbol{v}_{i+\delta}\neq \boldsymbol{0}.$ Therefore, $u_{mb}(z)$ is of the form
\begin{eqnarray*}
\boldsymbol{u}_{mb}(z)=\boldsymbol{u}'_{mb}(z)+\boldsymbol{u}''_{mb}(z) ~~~~~~~~~~~~~~~~~~~~~~~~~~\\
=\sum_{k=1}^{i-1}\boldsymbol{u}'_{mb,k}z^k + \sum_{k=i+\delta}^{\infty}\boldsymbol{u}''_{mb,k}z^k ~~~~~~~~~~~~\\
\boldsymbol{u}_{mb}(z)=\left[\boldsymbol{u}'_{mb,0},..,\boldsymbol{u}'_{mb,i-1},\boldsymbol{0},\boldsymbol{0},..\right]~~~~~~~~~~~~~~~~~~~~~~~~~\\
~~~~~~~~~~~~~~~~~+\left[\boldsymbol{0},..,\boldsymbol{0},\boldsymbol{u}''_{mb,i+\delta},\boldsymbol{u}''_{mb,i+\delta+1},..\right]~~~~~~~~~~~~~~
\end{eqnarray*}
i.e,
\[
\boldsymbol{u}_{mb}(z)=\left[\boldsymbol{u}_{mb,0},\boldsymbol{u}_{mb,1},...,\boldsymbol{u}_{mb,i},...,\boldsymbol{u}_{mb,i+\delta-1},\boldsymbol{u}_{mb,i+\delta},..\right]
\]
where $\boldsymbol{u}_{mb,i}=\boldsymbol{u}_{mb,i+1}=...=\boldsymbol{u}_{mb,i+\delta-1}=\boldsymbol{0} \in \mathbb{F}_q^b.$ 

With the minimal encoder $E_{mb}$, which has utmost $\nu_{b}$ memory elements, these $\delta$ consecutive zeros of would result in the state vector $\boldsymbol{\sigma}_{mb,t}$ becoming zero at time instant $i+\nu_{b}{\leq}i+\delta {\leq}j$, i.e, 
$\boldsymbol{\sigma}_{mb,i+\nu_{b}}=\boldsymbol{0}.$ But the definition of $S_{d_{free}}$ excludes such a condition, which means that $\boldsymbol{v}_{[0,j)} \notin S_{d_{free}},$ contradicting our original assumption. Thus we have proved our claim. 

\end{IEEEproof}

We shall now prove the following bound on $T_{d_{free}}({\cal C})$.
\begin{proposition}
\label{prop}
Let $\cal C$ be a $(c,b,\delta)$ convolutional code. Then
\begin{equation}
\label{eqn19}
T_{d_{free}}({\cal C}) \leq \left(d_{free}\left({\cal C}\right)-1\right)\delta+1.
\end{equation}
\end{proposition}
\begin{IEEEproof}
Let $\boldsymbol{v}_{[0,j)} \in S_{d_{free}}$ be some truncated codeword. Then we have $w_H\left(\boldsymbol{v}_{[0,j)}\right) \leq d_{free}\left({\cal C}\right)-1.$ By Lemma \ref{deltazeroes}, we have that in any consecutive $\delta$ segments, the Hamming weight of  $\boldsymbol{v}_{[0,j)}$ is at least $1.$ With this observation, and by the definition of $T_{d_{free}}({\cal C})$, we have (\ref{eqn19}), thus proving the proposition.
\end{IEEEproof}
Thus, for a network error correcting MDS convolutional code ${\cal C}_s$, we have the following bound on $T_{d_{free}}({\cal C}_s).$
\begin{corollary}
If the code ${\cal C}_s$ chosen in the construction of Subsection \ref{construction} is an $(n,k)$ MDS convolutional code, then we have the following bound on $T_{d_{free}}({\cal C}_s).$
\begin{equation}
\label{boundtdf}
T_{d_{free}}({\cal C}_s) \leq 6nk-2k^2+1.
\end{equation}
\end{corollary}
\begin{IEEEproof}
In the Construction of Subsection \ref{construction}, if the code ${\cal C}_s$ selected is an MDS convolutional code, then we know from the proof of Theorem \ref{fieldsizebound} that the degree being $\delta = 2k$ satisfies the required error correcting capability. Moreover, for an $(n,k,\delta)$ MDS convolutional code, we have 
\[
d_{free}({\cal C}) = (n-k)(\lfloor\delta/k\rfloor + 1) + \delta + 1
\]

Therefore, substituting this value for $d_{free}({\cal C}_s)$ with $\delta = 2k$ in (\ref{eqn19}) of Proposition \ref{prop}, we have (\ref{boundtdf}).
\end{IEEEproof}
\section{Illustrative Examples}
\label{sec5}
\subsection{Code construction for the butterfly network}
\begin{figure}[htbp]
\centering
\includegraphics[totalheight=2.2in,width=3.6in]{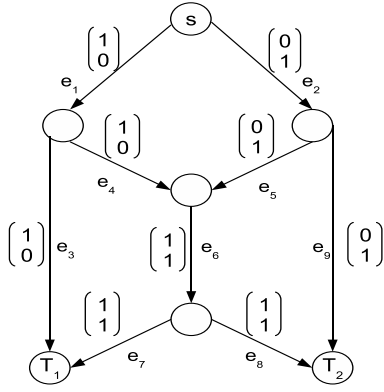}
\caption{Butterfly network}	
\label{fig:ButterflyNetwork}	
\end{figure}

The most common example quoted in network coding literature, the butterfly network, is shown in Fig. \ref{fig:ButterflyNetwork}. Let us assume the ancestral ordering as given in the figure. Every edge is assumed to have unit capacity. It is noted that the network code in place for the butterfly network as shown is a generic network code for all field sizes. We seek to design a convolutional code for this network which will correct all single edge errors.
\begin{example}[Butterfly network under a binary field]
The network transfer matrix for sink $T_1$ is the full rank $2 \times 2$ matrix
\[
M_{T_1} = \left[ \begin{array}{cc}
1 & 1 \\
0  & 1 \end{array} \right]=AF_{T_1}
\]
where 
\[
A =\left[ \begin{array}{ccccccccc}
1 & 0 & 0 & 0 & 0 & 0 & 0 & 0 & 0\\
0  & 1 & 0 & 0 & 0 & 0 & 0 & 0 & 0 \end{array} \right]
\]
\[
F_{T_1} = \left[ \begin{array}{ccccccccc}
1 & 0 & 1 & 0 & 0 & 0 & 0 & 0 & 0\\
1 & 1 & 0 & 1 & 1 & 1 & 1 & 0 & 0
\end{array} \right]^T
\]

Similarly, for sink $T_2$, $ M_{T_2} = \left[ \begin{array}{cc}
1 & 0 \\
1  & 1 \end{array} \right]=AF_{T_2}$ 
where 
\[
F_{T_2} = \left[ \begin{array}{ccccccccc}
1 & 1 & 0 & 1 & 1 & 1 & 0 & 1 & 0\\
0 & 1 & 0 & 0 & 0 & 0 & 0 & 0 & 1
\end{array} \right]^T
\]

For single edge errors, the error pattern set is 
\[
\Phi=\left\{\left\{e_i\right\}:i=1,2,...,9\right\}.
\]
Then the set of $9$ length error vectors over $\mathbb{F}_2$, ${\cal W}_{\Phi} = $
\[
\left\{(1,0,0,...,0),(0,1,0,...,0),...,(0,0,0,...,0,1),(0,0,0,...,0)\right\}
\]

For both sinks $T_1$ and $T_2$, we have 
\[
{\cal W}_{T}={\cal W}_{T_1} = {\cal W}_{T_2} = \left\{(0,0),(0,1),(1,0),(1,1)\right\}
\]
Since $M_{T_1}^{-1}=M_{T_1}$ and $M_{T_2}^{-1}=M_{T_2}$, we have 
\[
 {\cal W}_s =\bigcup_{T\in{\cal T}} \left\{\boldsymbol{w}_{_T}M_T\text{ }|\text{ }\boldsymbol{w}_{_T}\in{\cal W}_{T}\right\} 
\] 
\[
{\cal W}_s = \left\{(0,0),(0,1),(1,0),(1,1)\right\}
\]

Now we have $t_s = \max_{\boldsymbol{w}_s \in {\cal W}_s}w_H(\boldsymbol{w}_s) = 2.$ Hence a convolutional code with free distance at least $2t_s+1 = 5$ is required to correct these errors. With the min-cut $n$ being 2, let $k = 1$. Let this input convolutional code ${\cal C}_s$ be generated by the generator matrix $G_I(z) = \left[1 + z^2\text{ }\text{ }1+z+z^2\right].$

This code is a degree $2$ convolutional code with free distance $5$, and $T_{d_{free}}({\cal C}_s)=6$. Hence, by Theorem \ref{maintheorem}, this code will correct all single edge errors under the condition that consecutive single edge errors occur within $6$ network uses. Now the sinks must select between Case A and Case B for decoding, based upon their output convolutional codes.

The output convolutional code that is `seen' by the sink~$T_1$  has a generator matrix 
\[
G_{O,T_1}(z) = G_I(z)M_{T_1} = [1+z^2\text{ }\text{ }z].
\]
This code seen by sink $T_1$ has a free distance of only $3$, which is lesser than $2\left(\max_{\boldsymbol{w}_{T_1} \in {\cal W}_{T_1}} w_H(\boldsymbol{w}_{T_1})\right)+1 = 5$. Hence case B applies and decoding is done  on the trellis of the input convolutional code after processing. \\
Similarly, the convolutional code thus seen by the sink node $T_2$ has the generator matrix 
\[
G_{O,T_2}(z) = G_I(z)M_{T_2}(z) = [z\text{ }\text{ }1+z+z^2]
\]
This is a free distance $4$ code, which is again lesser than $2\left(\max_{\boldsymbol{w}_{T_2} \in {\cal W}_{T_2}} w_H(\boldsymbol{w}_{T_2})\right)+1 = 5$. Hence, for this sink too, Case B applies and decoding is done on the input trellis. \newline
\end{example}
\begin{example}[Butterfly network under a ternary field]
We now present another example to illustrate the case when the field size and the choice of the input convolutional code affects the error correction capabilities of the output convolutional codes at the sinks.
Let us assume the butterfly network with the network code being the same as the previous case, but over $\mathbb{F}_3$. The network transfer matrices in this case are the same as before, but the symbols are from $\mathbb{F}_3$.
We seek to correct single edge errors in this case too. Thus the error pattern set is the same as the previous case. Now we have the set of $9$ length error vectors over $\mathbb{F}_3$
\begin{eqnarray*}
{\cal W}_{\Phi} = \left\{(1,0,0,...,0),(0,1,0,...,0),...,(0,0,0,...,0,1),\right. \\ \left. (2,0,0,...,0),(0,2,0,...,0),...,(0,0,0,...,0,2),(0,0,0,...,0)\right\}
\end{eqnarray*}
The other sets are as follows. ${\cal W}_{T}={\cal W}_{T_1} = {\cal W}_{T_2} =$
\[
\left\{(0,0),(0,1),(1,0),(1,1),(0,2),(2,0),(2,2)\right\}
\] 
With $ M_{T_1}^{-1} = \left[ \begin{array}{cc} 1 & 2 \\ 0  & 1 \end{array} \right] 
\text{ and } 
M_{T_2}^{-1} = \left[ \begin{array}{cc}
1 & 0 \\
2  & 1 \end{array} \right], $
we have 
\[
{\cal W}_s = \bigcup_{T\in{\cal T}} \left\{\boldsymbol{w}_{_T}M_T^{-1}\text{ }|\text{ }\boldsymbol{w}_{_T}\in{\cal W}_{T}\right\}
\]
\[
=\left\{(0,0),(0,1),(1,0),(1,2),(0,2),(2,1),(2,0)\right\}
\]
Thus, again we have, 
\[
t_s = \max_{\boldsymbol{w}_s \in {\cal W}_s}w_H(\boldsymbol{w}_s) = \max_{\boldsymbol{w}_{_T} \in {\cal W}_T}w_H(\boldsymbol{w}_{_T}) = 2.
\]
Hence a convolutional code with free distance at least $2t_s+1 = 5$ is required to correct all single errors.

We compare the error correction capability of the output convolutional code at each sink for two input convolutional codes, ${\cal C}_s$ and ${\cal C}_s'$ generated by the matrices 
\[
G_I(z) = \left[1 +z^2\text{ }\text{ }1+z+z^2\right]
\]
and 
\[
G_I'(z) = \left[1 + z^2\text{ }\text{ }1+z+2z^2\right]
\]
respectively, each over $\mathbb{F}_3[z]$.
Both of these codes are degree $2$ convolutional codes and have free distances $d_{free}({\cal C}_s)=d_{free}({\cal C}_s')=5$, with $T_{d_{free}}({\cal C}_s)=T_{d_{free}}({\cal C}_s')=6$.

First, we discuss the case where the input convolutional code is ${\cal C}_s$. The sink $T_1$ thus sees the code generated by 
\[
G_{O,T_1}(z) = G_I(z)M_{T_1} = [1 + z^2\text{ }\text{ }2+z+2z^2]
\]
which has a free distance of $5$, with $T_{d_{free}}({\cal C}_{T_1}) = 6 = T_{d_{free}}({\cal C}_s)$. Thus decoding is done on the output trellis at sink $T_1$ to correct all single edge errors that as long as they are separated by $6$ network uses. Sink $T_2$ sees the code generated by 
\[
G_{O,T_2}(z)=[2+z+2z^2\text{ }\text{ }1+z+z^2]
\]
which has $d_{free}=6$, and $T_{d_{free}}({\cal C}_{T_2}) = 6 = T_{d_{free}}({\cal C}_s).$

Therefore, sink $T_2$ can also decode on the output trellis after multiplication by the corresponding processing matrix to get the required error correction in every $6$ network uses. Upon carrying out a similar analysis with the input convolutional code being ${\cal C}_s'$, we give the following tables for comparison. 
\begin{table}[htbp]
\centering
\caption{Butterfly network with ${\cal C}_s[d_{free}({\cal C}_s)=5,T_{d_{free}}({\cal C}_s)=6]$} \begin{tabular}{|c|c|c|c|}\hline
\textbf{Sink} & \textbf{Output convolutional } & \textbf{$d_{free}({\cal C}_{T_i})$}, & \textbf{Decoding on}\\
 & \textbf{code} $[G_{O,T_i}(z)]$ & $T_{d_{free}}({\cal C}_{T_i})$ & \\
\hline
$T_1$ & $[1+z^2\text{ }\text{ }2+z+2z^2]$ & 5,6 & Output trellis \\
\hline
$T_2$ & $[2+z+2z^2\text{ }\text{ }1+z+z^2]$ & 6,6 & Output trellis\\
\hline
\end{tabular}
\label{tab3}
\end{table}
\begin{table}[htbp]
\centering
\caption{Butterfly network with ${\cal C}_s' [d_{free}({\cal C}_s')=5,T_{d_{free}}({\cal C}_s')=6 ]$} \begin{tabular}{|c|c|c|c|}\hline
\textbf{Sink} & \textbf{Output convolutional} & \textbf{$d_{free}({\cal C}_{T_i})$}, & \textbf{Decoding on}\\
 & \textbf{code} $[G_{O,T_i}(z)]$ & $T_{d_{free}}({\cal C}_{T_i})$ & \\
\hline
$T_1$ & $[1+z^2\text{ }\text{ }2+z]$ & 4,3 & Input trellis \\
\hline
$T_2$ & $[2+z\text{ }\text{ }1+z+2z^2]$ & 5,5 & Output trellis\\
\hline
\end{tabular}
\label{tab4}
\end{table}
With the input convolutional code being ${\cal C}_s$, conditions (\ref{decodAcond1}) and (\ref{decodAcond2}) are satisfied at both sinks. Hence additional processing can be avoided at both sinks and they can decode on the output trellis directly, and get single edge error correction under the constraint that consecutive single edge errors are separated by at least $6$ network uses.

However with ${\cal C}_s'$, one of the sinks $T_1$ does not have sufficient free distance at its output convolutional code, and hence has to process the incoming symbols using $M_{T_1}^{-1}$ and thereby decode on the trellis of the input convolutional code. 

Thus it can be seen that using a larger field size and choosing the input convolutional code appropriately can give more desirable properties to the output convolutional codes at the sinks.
\end{example}
\subsection{Code construction for the $_4C_2$ network}
\begin{example}[$_4C_2$ combination network under $\mathbb{F}_3$]
\label{combiexample1}
Let us consider the combination network in Fig. \ref{fig:4C2Network}. The network transfer matrices for the $6$ sinks are as in Table \ref{tab1}. We seek to design a convolutional code that will correct all network-errors whose error vectors have Hamming weight utmost 2 (i.e single and double edge errors).

The error pattern set is thus 
\[
\Phi~=~\left\{\left\{e_i,e_j\right\}:i,j=1,2,...,15,16 \text{ and }i \neq j\right\}
\]
The set ${\cal W}_{\Phi}$ is the set of all $16$ length vectors with Hamming weight utmost $2.$ We have 
\[
{\cal W}_{T_1} = {\cal W}_{T_2} = ... = {\cal W}_{T_6} = \mathbb{F}^2_3 
\]
and 
\[
{\cal W}_s = \bigcup_{T\in{\cal T}} \left\{\boldsymbol{w}_{_T}M_T^{-1}\text{ }|\text{ }\boldsymbol{w}_{_T}\in{\cal W}_{T}\right\} = \mathbb{F}^2_3
\]
For every sink $T_i$, we have 
\[
\max_{\boldsymbol{w}_{T_i} \in {\cal W}_{T_i}}w_H(\boldsymbol{w}_{T_i})= \max_{\boldsymbol{w}_s \in {\cal W}_s}w_H(\boldsymbol{w}_s)= t_s = 2
\]
Therefore the input convolutional code needs to have free distance at least $5$.

As in Example \ref{exm1}, let the input convolutional code, ${\cal C}_{s}$, over $\mathbb{F}_3[z]$ be generated by the matrix 
\[
G_I(z) = \left[1 + z^2\text{ }\text{ }1+z+z^2\right].
\]
This code has free distance = $5$, and $T_{d_{free}}({\cal C}_{s})=6.$

Each sink decodes on either the input or the output trellis depending upon whether $d_{free}({\cal C}_{T_i}) \geq 2t_s+1$, and if $ T_{d_{free}}({\cal C}_{s}) \geq T_{d_{free}}({\cal C}_{T_i})$, and hence can correct all network-errors with with their pattern in $\Phi$ as long as consecutive errors are separated by $6$ network uses. The output convolutional codes at the sinks, their free distances and their $T_{d_{free}}({\cal C}_{T_i})$ are shown in in Table \ref{tab2}.

\begin{table}[htbp]
\centering
\caption{$_4C_2$ network with $G_I(z) = \left[1 + z^2\text{ }\text{ }1+z+z^2\right]$} 
\begin{tabular}{|c|c|c|c|}\hline
\textbf{Sink} & \textbf{Output} & \textbf{$d_{free}({\cal C}_{T_i})$}, & \textbf{Decoding on}\\
 & \textbf{convolutional code} & $T_{d_{free}}({\cal C}_{T_i})$ & \\
\hline
$T_1$ & $[1+z^2\text{ }\text{ }1+z+z^2]$ & 5,6 & Output trellis\\
\hline
$T_2$ & $[1+z^2\text{ }\text{ }2+z+2z^2]$ & 5,6 & Output trellis \\
\hline
$T_3$ & $[1+z^2\text{ }\text{ }2z]$ & 3,4 & Input trellis \\
\hline
$T_4$ & $[1+z+z^2\text{ }\text{ }2+z+2z^2]$ & 6,6 & Output trellis\\
\hline
$T_5$ & $[1+z+z^2\text{ }\text{ }2z]$& 4,5 & Input trellis\\
\hline
$T_6$ & $[2+z+2z^2\text{ }\text{ }2z]$& 4,5 & Input trellis\\
\hline
\end{tabular}
\label{tab2}
\end{table}
\end{example}
\section{Comparison with block network error correction codes}
\label{sec5.5}

The approach of \cite{YaY} can also be used to obtain network error correcting codes that correct $t$ edge errors once in every $J$ network uses (for some positive integer $J$). A time-expanded graph would then be used, i.e, with the network nodes (those except the source and sinks) and edges replicated for each additional time instant. 

Suppose the network has been replicated $J$ times. Then the algorithm in \cite{YaY} can be employed to obtain a $t$-error correcting BNECC for the time-expanded network, which equivalently for the original network gives a network error correcting code that corrects $t$ errors once in every $J$ network uses. It is noted that the sufficient field size $q$ required by the technique of \cite{YaY} to construct a $t$-error correcting BNECC for the time-expanded graph ($\cal T$ being the set of all sinks) is such that 
\[
q > \sum_{T\in\cal{T}}
\left(
\begin{array}{c}
J|\cal{E}| \\
2t
\end{array}
\right).
\]

Our approach demands a field size according to Theorem \ref{fieldsizebound}, which is independent of the number of edges in the network. Although the error correcting capability might not be comparable to that offered by the BNECC, the reduction in field size is a considerable advantage in terms of the computation to be performed at each coding node of the network. Also, the use of convolutional codes permits decoding using the Viterbi decoder, which is readily available. 

\begin{figure}[htbp]
\centering
\includegraphics[totalheight=2.8in,width=3.6in]{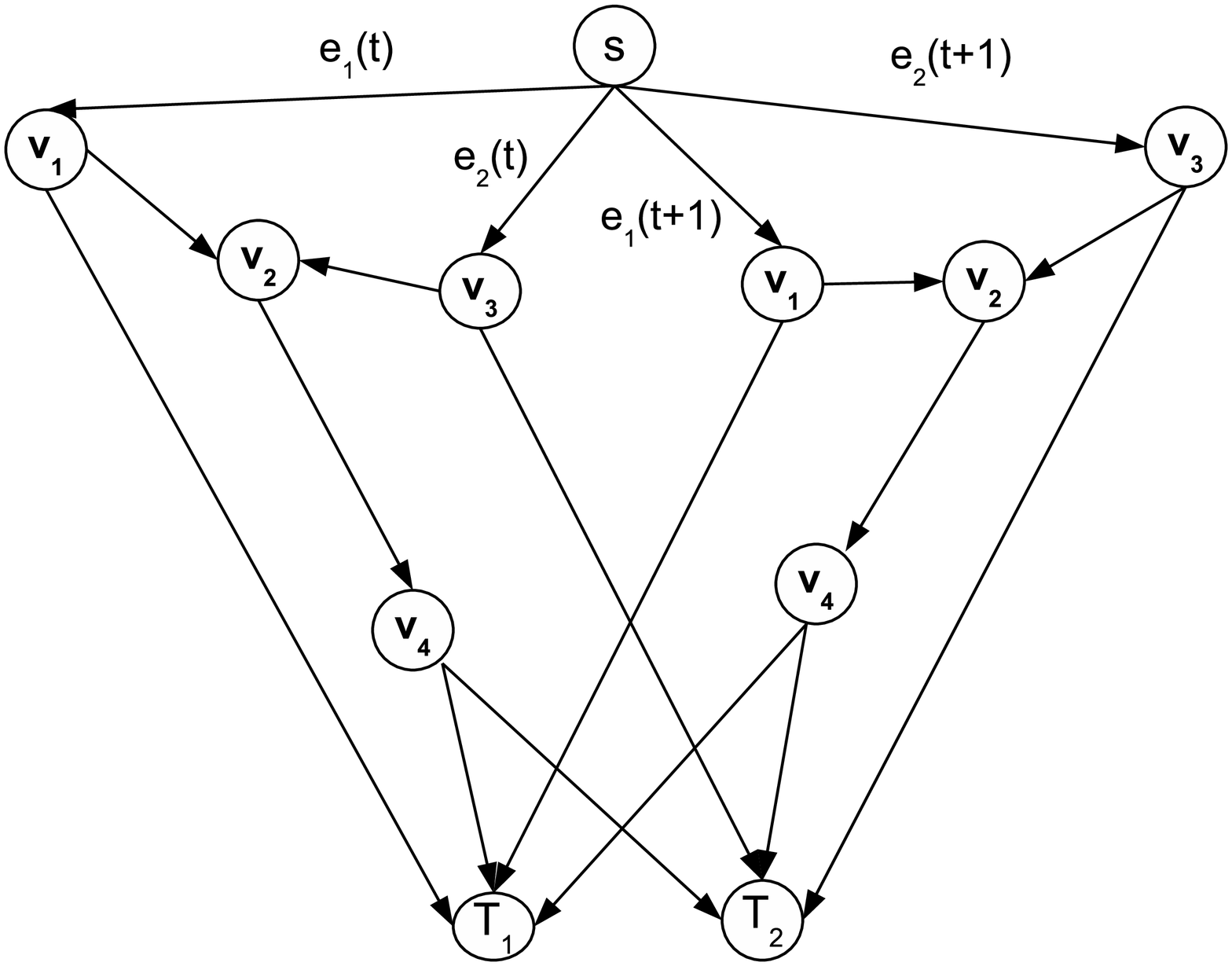}
\caption{A twice replicated butterfly network. The edges are marked with a time index as to denote the time-expanded nature of the network.}	
\label{fig:butterflytwice}	
\end{figure}

For example, one could design network error correcting codes according to \cite{YaY} for the butterfly network by using the twice replicated butterfly network as shown in Fig. \ref{fig:butterflytwice}. The time-expanded network has min-cut 4, and thus the technique in \cite{YaY} can be used to obtain BNECCs, which correct single or double edge errors in the butterfly network once in $2$ network uses. In either case, the sufficient field size $q$ is such that $q > 306$, although by trial and error a code could be found over a smaller field size. On the other hand, the convolutional code that we used here in our paper for the butterfly network is over the binary and ternary fields.

\section{Discussion}
\label{sec6}

In the construction of Subsection \ref{construction}, the maximum Hamming weight $t_s$ of the vectors in the set ${\cal W}_s$, is such that $t_s\leq n.$ Clearly the actual value of $t_s$ is governed by the network code and hence the network code influences the choice of the network-error correcting convolutional code. Therefore the network code designed should be such that $t_s$ is minimal, so that the free distance demanded of the network-error correcting convolutional code in the construction of Subsection \ref{construction} is minimal.

Also, for a particular error pattern set, the decoding procedure at the sinks (Case-A or Case-B of decoding as in Subsection \ref{decoding}) is influenced by the field size, the network code and the network-error correcting convolutional code chosen. The examples given in Section \ref{sec5} illustrate the construction of Subsection \ref{construction} and also compare the effects of change in field size and the convolutional code chosen to correct errors corresponding to a given fixed error pattern set. 

%
\section*{Acknowledgment} This work was supported  partly by the DRDO-IISc program on Advanced Research in Mathematical Engineering through a research grant to B.~S.~Rajan.

\end{document}